# A Model-Based Architecture for Automatic Anesthesia Infusion

Liu Da-yang

(School of Computers, Guangdong University of Technology, Guangzhou 510006, China)

**Abstract**

This paper presents a model-based control architecture. Based on the Medical Cyber-physical Systems (MCPS) concept, we construct a safe and reliable automatic anesthesia control closed-loop system. The control architecture uses the patient's propofol pharmacokinetic/pharmacodynamic model, using the calculated results as a feedback signal to correct the model's uncertainties. During anesthesia, the bispectral index (BIS) level is used as a control variable to control the depth of anesthesia. The proposed control architecture can automatically adjust the infusion rate of anesthesia drugs without concussion or overshoot. The system adaptively adjusts the drug infusion rate to maintain the BIS level at a stable target value. We utilized MATLAB-Simulink to verify the effectiveness of this method, and the robustness test was carried out by adding noise blocks. At last, we analyze the simulation result of the average patient. In general, the proposed method achieves a stable and fast induction period during the maintenance phase of anesthesia with a good disturbance rejection.

## 1. Introduction

Anesthesia management is prone to mistakes. During the period of anesthesia, the anesthetist will adjust the dose of anesthetics according to the operation situation and the patient's vital signs[1]. However, due to the complexity of the clinical environment and the patient's differences, even experienced anesthetists are not immune to cardiovascular or neurological damage due to over administration of drugs or inadequate administration of drugs leading to intraoperative awareness. Ira Dhawan[2] counted the reports of relevant anesthesia cases from 1998 to 2011 and pointed out that the incidence of drug administration errors was 0.078%-0.75%, with the two most significant categories of errors involving incorrect doses (20%) and incorrect drugs (20%). Infusion pumps and other systems have not established effective feedback mechanisms with patient monitoring systems, which is also a problem to be solved urgently. We hope that there will be a way to achieve some degree of improvement in terms of anesthesia optimization and patient risk reduction.

The 4th Industrial Revolution has brought a new expansion called Health 4.0[3], in which Cyber-physical Systems (CPS) are the core technology. CPS integrates the physical world and the information world, forming a new structure combining hardware and software and becoming the core technology system to support and lead the transformation of the new generation of industry[4]. Medical cyber-physical systems (MCPS) are healthcare critical integration of a network of medical devices[5]. MCPS regards the traditional clinical scenario as a closed-loop system. The nursing staff is the controller, the medical equipment is the sensor and actuator, and the patient is the control object[6]. Analyzing healthcare-related data captured in real-time enables precise patient care, and context-aware approaches guide expert decision-making.

Automatic intravenous anesthesia, consisting of monitoring systems, control software, and communication channels, delivers preferable effects than artificial operation and Target Controlled Infusion (TCI)[7]. Although automated anesthesia systems are a hot topic, few studies have reported their implementation in MCPS. Control integration of patient care is still lacking in healthcare and most anesthesia fields. In reality, we are unlikely to test directly on living patients, so early validation based on model simulation is needed[7]. It is easier and less costly to make changes at the model level.

The structure of the paper is as follows. The following section introduces the concepts of open and closed-loop drug delivery systems in clinical anesthesia practice. The third section offers the patient model in anesthesia control systems. The overview of the control strategy is referred to in the fourth section, while the result and analysis are addressed in the final section.

## 2. Introduction to automatic anesthesia system

### 2.1 Open Loop and Closed Loop in Anesthesia

Anesthesia includes three parts: unconsciousness, analgesia, and neuromuscular block (NMB)[8]. The introduction of monitoring technologies such as depth of consciousness monitor can timely detect complications and alarm, providing safe and predictable anesthesia and paving the way for the research of automatic closed-loop anesthesia systems. Fig.1 shows the anesthesia depth monitor equipment, hemodynamics, respiratory dynamics, etc., recommended in the anesthesia standard[8]. Anesthetists complement clinical observations to maintain the optimal infusion rate during anesthesia. The context-aware control system could free the anesthetist from repetitive assignments, allowing them to focus on other high-level tasks. In addition, it is necessary to address the particular hazardous factor associated with infected patients caused by the COVID-19 outbreak[9].

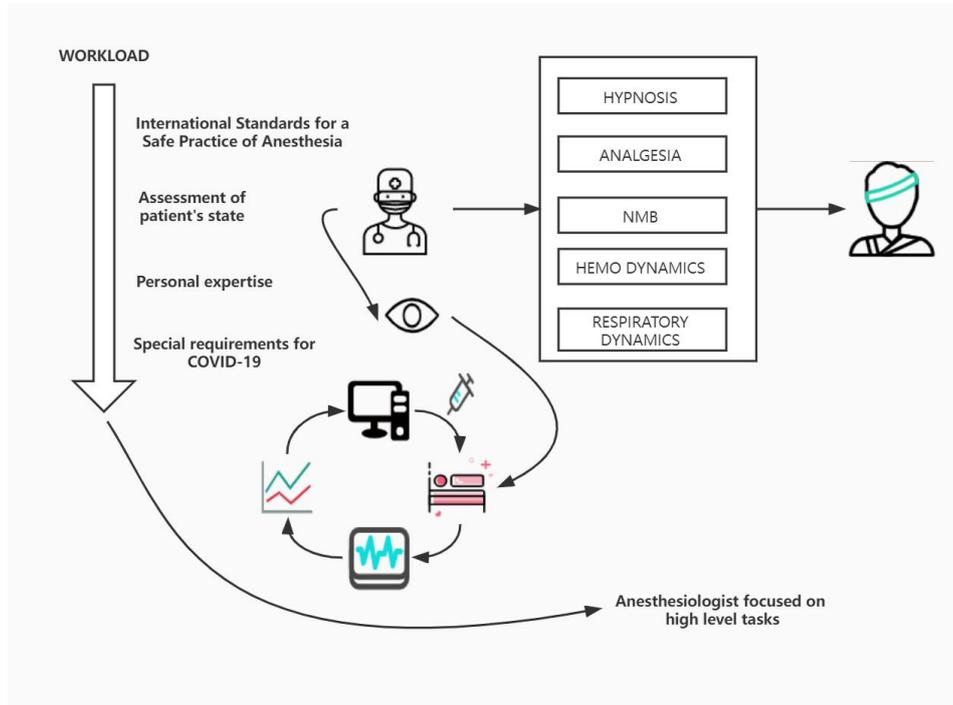

Fig.1. Clinical nursing model of general anesthesia patients' condition monitoring.

At present, the computer mainly controls the drug infusion through the open-loop TCI system to control a stable anesthesia drug delivery rate. The TCI system primarily uses the pharmacokinetics or the plasma concentration prescribed by medical practitioners to simulate the aimed effect-site[10]. Nevertheless, the anesthetist just simply selects the initial target dose or intervention and then adjusts it by assessing the status of patients during surgery. Clinicians hypothesize the drug's impact on the patient according to monitoring results or experience. In contrast, the proposed closed-loop control system uses direct measurements from the anesthesia depth monitor to automatically regulate the infusion rate. It is much more reliable than the subjective judgment of the anesthetist.

It is unnecessary to categorize the anesthesia system into separate groups in a practical clinical setting. The solution can be to use a hybrid system that allows the controller to respond to the anesthetist's movements to avoid complications[11]. The system can support the anesthetist in clinical decision-making by reasoning, optimizing drug administration, and minimizing accident overdosage or under dosage. The technology could also bring professional guidance to under-resourced areas.

**2.2 Control System for Anesthesia Application**

The integrated closed-loop system is shown in Fig.2. It integrates hand-operated clinical practice and automated control to achieve optimal drug infusion. The system adopts a multiple-input-multiple-output (MIMO) scheme, which considers common anesthetic drugs in clinical practice and their influence on individual output to avoid omission. In the system, an infusion pump is used to infuse various medications. The sensors relay information about the patient's physiological dynamics. Patients are systems with complex dynamics. The anesthetist can inject the patient directly when needed, creating interference in the control system. The controller estimates and adjusts the drug dosage based on the patient's current status. The model should be adaptive and optimized online to ensure real-time performance. Patient safety is the most crucial consideration in the system.

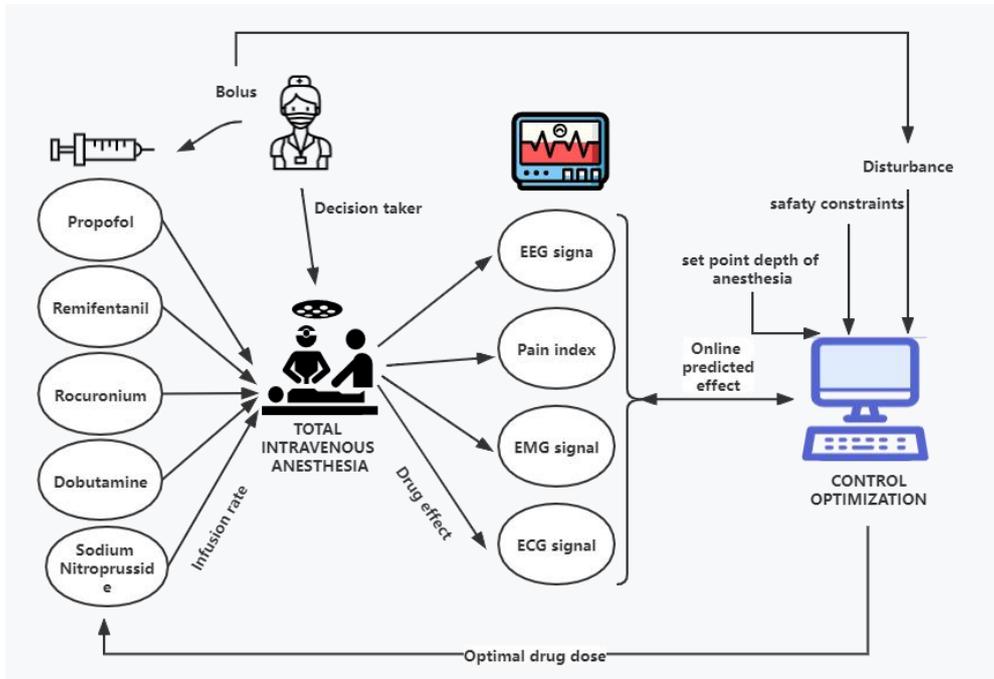

Fig.2. Integrated closed-loop system for general anesthesia automation.

Fig.3 shows the infusion pump drive module for cyber-physical interaction (injecting drugs into the human body). The control system uses the output of the pharmacokinetic model as a feedback signal to maintain the level of drug concentration in the blood to be the same as the reference value. The control algorithm of the system takes the reference value and the predicted value together as input. It calculates the required infusion rate through the error of the two and predicts the future drug concentration level.

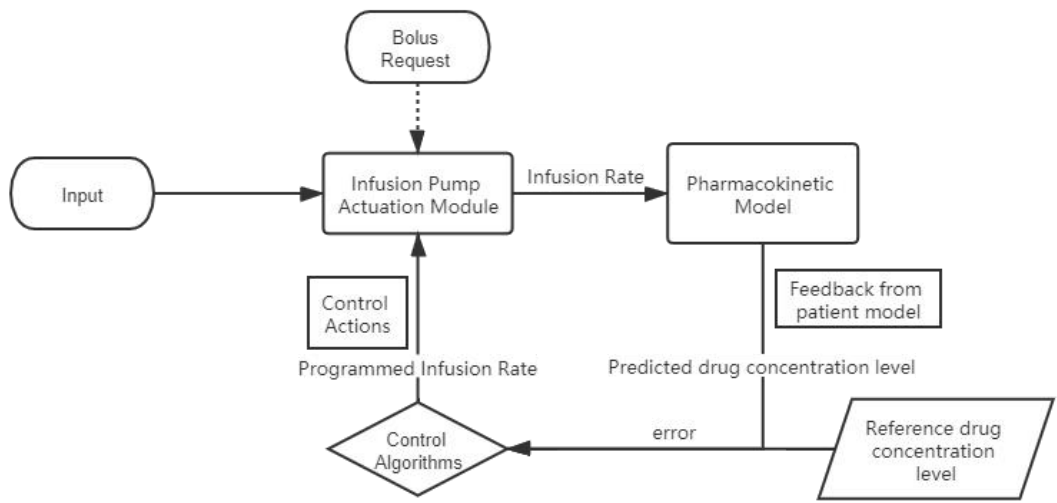

Fig.3. Infusion pump control system.

## 3. Pharmacokinetic-Pharmacodynamic modeling of patient

High variability is the characteristic of patient modeling in the anesthesia infusion system. Pharmacokinetic (PK) - pharmacodynamic (PD) compartment model is generally used to predict drug effect, and drug effect is measured by nonlinear Hill curve[12]. The pharmacokinetics model explains the association between drug infusion and drug concentration in vivo, and pharmacodynamics is the effect of the central compartment drug concentration on the patient's body.

Electroencephalogram (EEG) indicators could track Hypnosis-related changes caused by different drug concentrations. The Bispectral Index (BIS) is derived from EEG[13], and it is a more accurate method to judge the sedation level and monitor the depth of anesthesia. When BIS is 0, there is no EEG activity at all. BIS value of 100 represents a fully awake state, 60-70 and 40-60 represent calm and modest anesthesia status, respectively[14].

Propofol is a common intravenous anesthetic, and the following formula[15] describe the PK-PD mathematical model of propofol:

$$\begin{aligned}
\dot{c}_1(t) &= -[k_{10} + k_{12} + k_{13}] \cdot c_1(t) + k_{21} \cdot c_2(t) + k_{31} \cdot c_3(t) + u(t)/V_1 \\
\dot{c}_2(t) &= k_{12} \cdot c_1(t) - k_{21} \cdot c_2(t) \\
\dot{c}_3(t) &= k_{13} \cdot c_1(t) - k_{31} \cdot c_3(t) \\
\dot{c}_e(t) &= k_{1e} \cdot c_1(t) - k_{e0} \cdot c_e(t)
\end{aligned} \qquad (1)$$

Eq.(1) is the PK model of propofol, which uses the three-compartment model to denote the drug kinetic characteristics. The three compartments include a central compartment equivalent to blood and two peripheral compartments with different intake and release rates. In the formula, $c_1$ denotes the concentration of drug in the central compartment (representing plasma). The peripheral compartment with a faster rate of exchange with the central compartment is called the shallow compartment(e.g., muscles), and the peripheral compartment with a slow rate of drug exchange with the central compartment is called the deep compartment(e.g., fat). Their drug concentration is characterized by $c_2$ and $c_3$, respectively. U(t)[mg/min] is the infusion velocity of drugs into the central compartment. $V_1$ is the volume of compartment 1, 4.27[L]. $C_e$ is concentration in a hypothetical effect-site compartment, which is the site of drug action. As plasma is not the site of drug action, Schnider[19] et al. proposed the concept of effect compartment to explain the clinical phenomenon that the peak effect lags behind the peak plasma concentration. In clinical practice, the frequency drug transfer of propofol from the central compartment to the effect-site compartment is identified as the frequency of drug elimination from the effect-site compartment: $k_{e0}=k_{1e}=0.456[min^{-1}]$[16]. Parameter $k_{ij}$ (i ≠ j) represents the frequency of drug shift from compartment i from compartment j (ratio of the gap to volume). $K_{ij}$ relies on the patient's age, height, weight, and gender. The calculation method is shown as follow:

$$V_1 = 4.27[l], V2 = 18.9 - 0.391 \cdot (age - 53)[l], V_3 = 2.38[l]$$
$$C_{l1} = 1.89 + 0.456 \cdot (weight - 77) - 0.0681 \cdot (lbm - 59) + 0.264 \cdot (height - 177)[l/min]$$
$$C_{l2} = 1.29 - 0.024 \cdot (age - 53), C_{l3} = 0.836 \quad (2)$$
$$k_{10} = \frac{C_{l1}}{V_1}[min^{-1}], k_{12} = \frac{C_{l2}}{V_1}[min^{-1}], k_{13} = \frac{C_{l3}}{V_1}[min^{-1}]$$
$$k_{21} = \frac{C_{l2}}{V_2}[min^{-1}], k_{31} = \frac{C_{l3}}{V_3}[min^{-1}]$$

$C_{l1}$ is the velocity of the drug removed from the human body and needs to be calculated using lean body mass(LBM). $C_{l2}$ depends on age, and $C_{l3}$ is a constant. They are the velocities of drug transfer from the middle compartment to the other two compartments through diffusion. Eq.(3) represents the LBM for male (M) and female (F):

$$lbm\_m = 1.1 \cdot weight - 128 \cdot \frac{weight^2}{height^2}$$
$$lbm\_f = 1.07 \cdot weight - 148 \cdot \frac{weight^2}{height^2} \quad (3)$$

Eq.(4) is the PD model, also known as the Hill curve, which shows the correlation between BIS and concentration of effect-site $C_e$[17]. $E_0$ represents the waking state without drug infusion. $E_{max}$ represents the maximum effect achieved during infusion. $C_{e50}$ is the drug concentration at 50% of the maximal effect and means the patient's sensibility to the drug. γ determines the steepness of the curve.

$$BIS(t) = E_0 - E_{max} \cdot \frac{C_e(t)^\gamma}{C_e(t)^\gamma + C_{e50}^\gamma} \quad (4)$$

We selected the patient parameters used in [15] to assess the model's performance, and Table 1 shows the parameters. The 13<sup>th</sup> patient was a fictitious individual whose parameters were calculated from the algebraic average of other patient parameters.

Table 1 Parameters of patients

| Id | Age | Height[cm] | Weight[kg] | Sex | $C_{e50}$ | γ | $E_0$ | $E_{max}$ |
|----|-----|------------|------------|-----|-----------|-----|-------|-----------|
| 1  | 40  | 163        | 54         | F   | 6.33      | 2.24 | 98.8 | 94.10 |
| 2  | 36  | 163        | 50         | F   | 6.76      | 4.29 | 98.6 | 86.00 |
| 3  | 28  | 164        | 52         | F   | 8.44      | 4.10 | 91.2 | 80.70 |
| 4  | 50  | 163        | 83         | F   | 6.44      | 2.18 | 95.9 | 102.00 |
| 5  | 28  | 164        | 60         | M   | 4.93      | 2.46 | 94.7 | 85.30 |
| 6  | 43  | 163        | 59         | F   | 12.00     | 2.42 | 90.2 | 147.00 |
| 7  | 37  | 187        | 75         | M   | 8.02      | 2.10 | 92.0 | 104.00 |
| 8  | 38  | 174        | 80         | F   | 6.56      | 4.12 | 95.5 | 76.40 |
| 9  | 41  | 170        | 70         | F   | 6.15      | 6.89 | 89.2 | 63.80 |
| 10 | 37  | 167        | 58         | F   | 13.70     | 1.65 | 83.1 | 151.00 |

| 11 | 42 | 179 | 78 | M | 4.82 | 1.85 | 91.8 | 77.90 |
| 12 | 34 | 172 | 58 | F | 4.95 | 1.84 | 96.2 | 90.80 |
| 13 | 38 | 169 | 65 | F | 7.42 | 3.00 | 93.1 | 96.58 |

Fig.4 is a simulation of all 13 patients, showing the correlation between BIS output and effect-site compartment concentration. In clinical practice, the BIS level between 40 and 60 indicates the patient is in a better state of anesthesia, and the corresponding effect-site concentration is around 3 to 9.

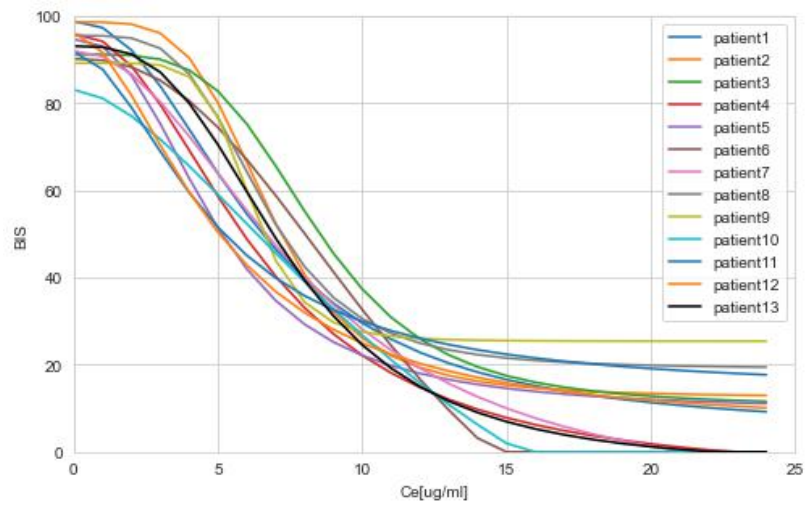

Fig.4. Simulation result of all 13 patients.

## 4. Control architecture design

According to the content depicted in the previous section, we propose a control architecture to transition the depth of anesthesia to the target BIS level and keep the BIS level at that stage. The control protocol sets the desired BIS at first. A BIS value between 40-60 is usually optimal in clinical practice, so we fix the BIS level at 50. Fig.5 shows the control architecture, where d(t) is the deviation between the target and calculated concentrations. The control action u(t) is the infusion speed of propofol, and the result of the patient linear model is the estimated effect-site concentration $C_e(t)$ of the patient. The output y(t) is the BIS level with noise interference added. The innovation signal i(t) is used to complement the deviation between the calculated result and the target value. We add noise blocks to represent the common BIS signal noise to verify the scheme's robustness. Therefore, we added two filters to the control scheme to deal with noise.

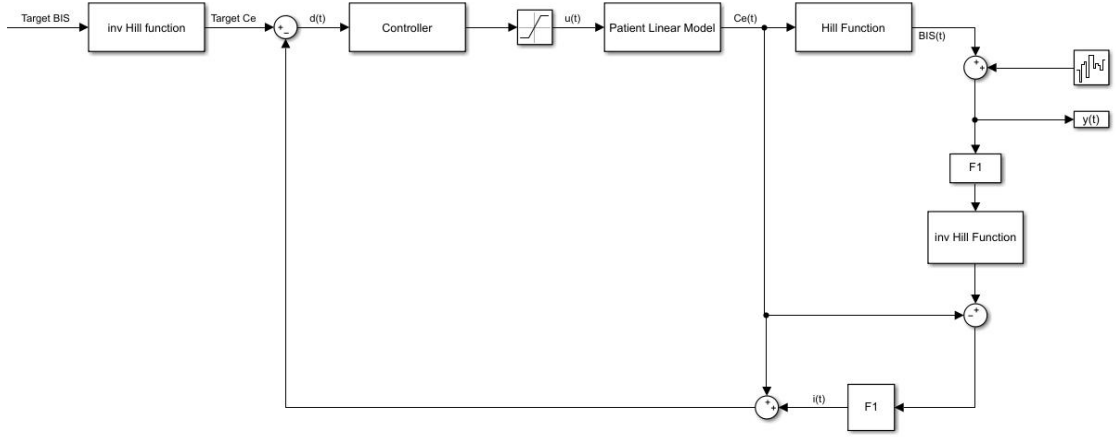

Fig.5. The control architecture for the automated regulation of propofol for anesthesia.

The architecture we put forward can integrate the patient model into the control architecture to provide exceptional and personalized drug infusion. The calculation result is used as a feedback signal to obtain an appropriate drug infusion rate. Other parameters ($C_{e50}$, $\gamma$, $E_{max}$) are usually unknown except $E_0$ can be measured when calculating BIS level, so we use the average parameters in Table 2, which are derived from[18]. Eq.(5) expresses the inverse of the Hill function in the control architecture.

$$C_e(t) = C_{e50}(t) \cdot \left( \frac{E_0 - BIS(t)}{E_{max} - E_0 + BIS(t)} \right)^{\frac{1}{\gamma}} \qquad (5)$$

Table 2  The average value of the parameters of the Hill function

| Parameter | Value |
|---|---|
| $E_{max}$ | 87.5 |
| $\gamma$ | 2.69 |
| $C_{e50}$ | 4.92 |

To study the robustness of the controller, we added mutational interference to the BIS measurement to simulate external nociceptive stimuli. Noise can degrade the property of the architecture; for instance, it may cause a risk of slow or unstable responses. To solve the noise problem, we use two linear filters. The $F_1$ filter pre-filters the BIS signal with the primary purpose of eliminating the potential disturbance with no free of affecting the system. The $F_2$ filter deal with the tradeoff between system and noise filtering for innovative signal i(t). We use second-order low-pass filters for both filters:

$$F_1(s) = \frac{1}{(T_{f1}s+1)^2}$$
$$F_2(s) = \frac{1}{(T_{f2}s+1)^2} \qquad (6)$$

Particularly, the integral absolute error (IAE) is taken as the fitness function:

$$IAE = \int_0^\infty |\text{target}BIS(t) - BIS(t)|\, dt. \quad (7)$$

The filter time constant $T_{f1}$ is fixed at 0.1 empirically. We use the gain scheduling method[17] to tune filter $F_2$. For $T_{f2}$, we calculated the IAE of each patient in Table.1 and compared it with the condition in the absence of disturbance to obtain the performance degradation ratio in the worst case:

$$d = \max_n \frac{IAE_{n,\text{filter}} - IAE_n}{IAE_n} \quad (8)$$

In the formula, n represents the nth patient. $IAE_{n,\text{filter}}$ represents the integral absolute error calculated with $T_{f2}$. $IAE_n$ represents the value computed without a filter in the absence of noise. Fig.6 shows the result of tuning the filter in the sustaining stage. A proper performance degradation ratio index can be designated at 30%, corresponding to the balance between noise filtering and optimal regulation. Finally, we obtain $T_{f2}$=9.7871 corresponding to disturbance rejection.

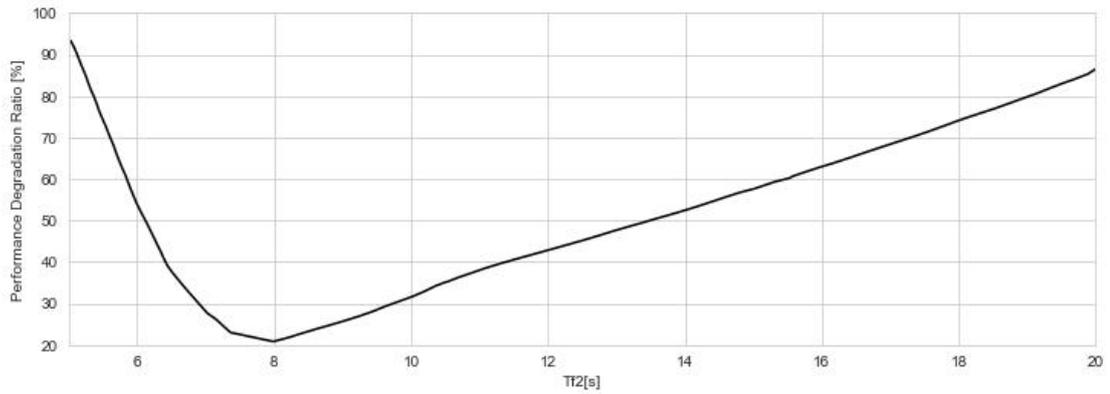

Fig.6. Performance degradation ratio to $T_{f2}$.

## 5. Simulation result and discussion

We simulated the proposed architecture in MATLAB-Simulink. In this section, we will analyze and discuss the simulation results. For observation purposes, we analyze the results of the 13th patient first to illustrate the reliability of the proposed system. The purpose of anesthesia induction is to achieve the desired depth of anesthesia as quickly as possible, usually within 4 minutes. We focused on how long it took patients to reach the target BIS value for the first time and how stable the injection rate was to maintain the BIS level.

### 5.1 Simulation result of the 13th patient

The simulation result of the 13th patient are shown in Fig.7:

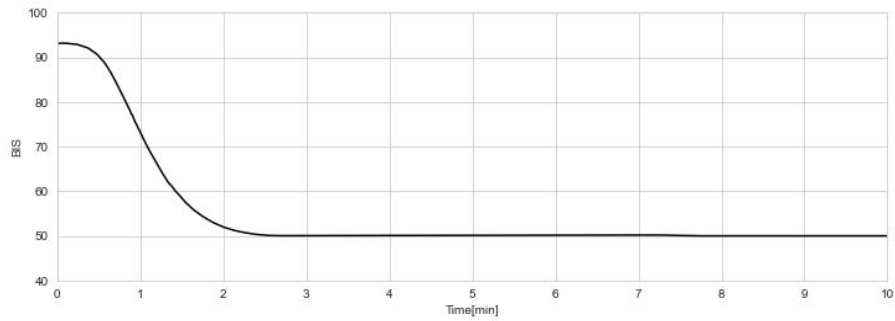

（a）The curve of BIS

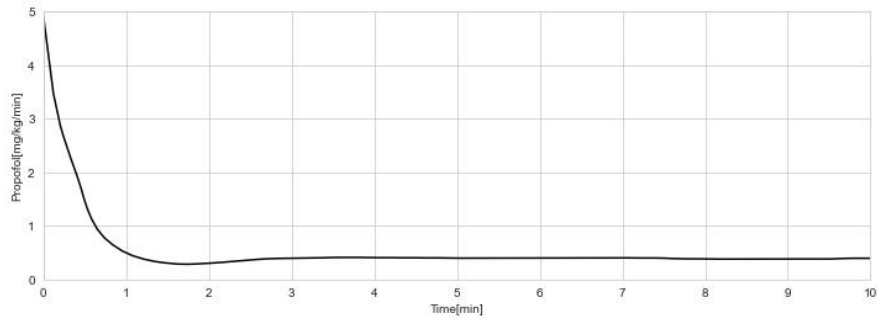

（b）The curve of injection rate

Fig.7.The response curve of the control system when BIS=50.

From a clinical point of perspective, the result is satisfactory. The BIS level of the simulation results reached the set value without the occurrence of undershoot, and the drug injection rate did not appear concussion or overshoot. After getting the preset value, it could remain stable all the time. Other signals obtained by model simulation are shown in the figure below:

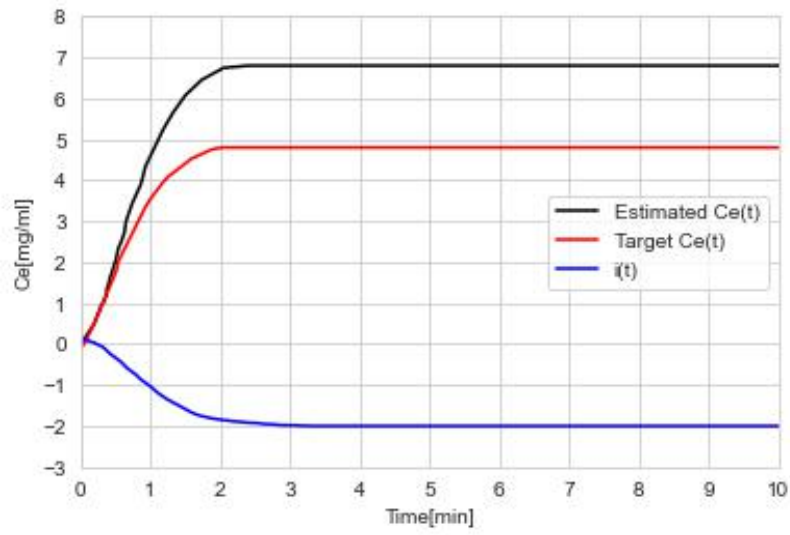

Fig.8.Model simulation response of the patient 13

Target $C_e(t)$ is a superior value figured by the inverse of the Hill function. Estimated $C_e(t)$ is the estimated effect-site concentration, also the real effect-site. The curve of i(t) is used as an additional corrective action to correct possible errors in the calculation so that the stable target BIS level can be achieved.

### 5.2 Test of Controller Robustness

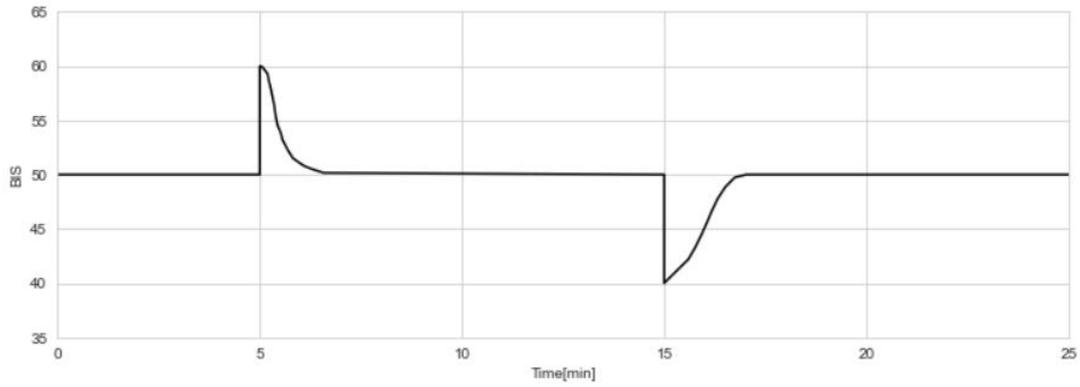

(a) BIS level

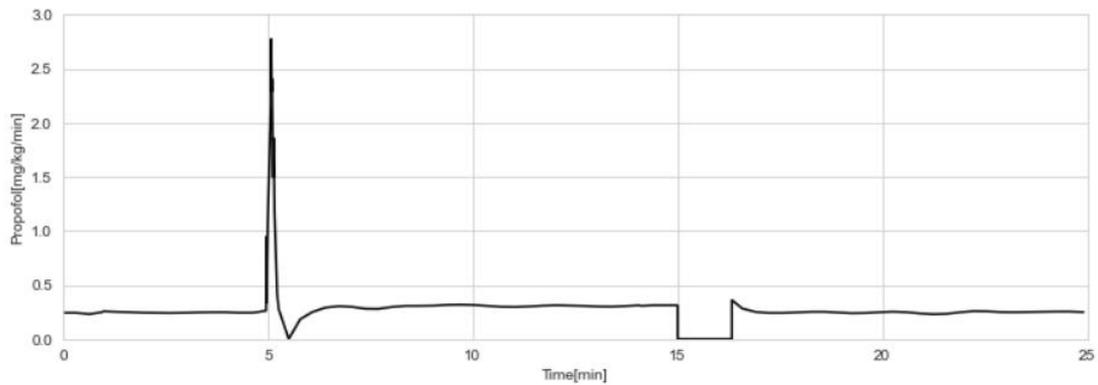

(b) Control action

Fig.9. Disturbance response for the 13th patient.

Fig.9 (a) shows the disturbance encountered while maintaining the BIS level, and (b) shows the corresponding control actions. When the BIS level rises, control action increases the drug infusion rate to reduce the patient's BIS level to avoid the occurrence of intraoperative awareness. On the contrary, when the BIS level drops, the control action will lessen the injection volume in time to ensure patient safety. Control action will maintain a stable infusion rate after adjusting the infusion volume and BIS level to the desired level.

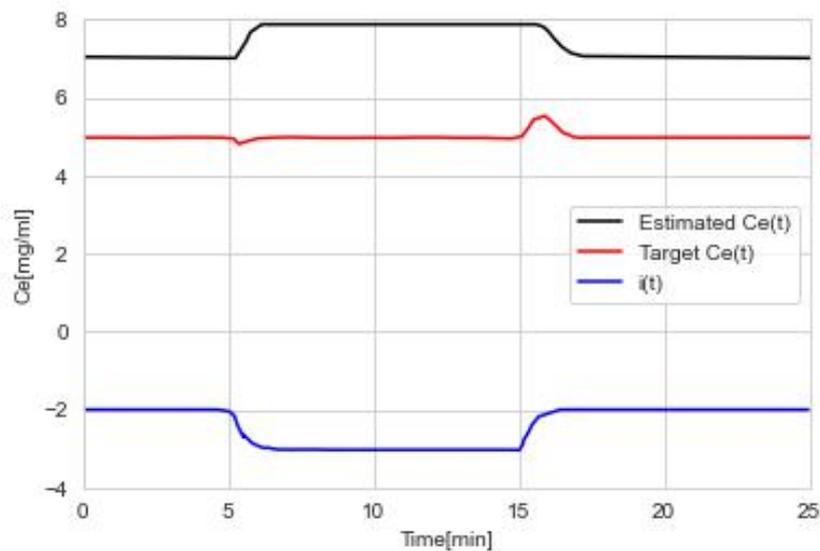

Fig.10. Disturbance rejection of the 13<sup>th</sup> patient.

Fig.10 draws the simulation result of the control architecture under the condition of disturbance rejection. The result shows that the interference has only a slight influence on the output process. The zero steady-state error can be realized by innovation signal i(t).

## 6. Conclusion

A model-based automatic control architecture was brought up for anesthesia. The system integrates the patient model so that the proposed control architecture could design a proper dosing regimen for each patient. The simulation result of a proper patient was mainly presented to illustrate the rationality of the proposed architecture. The robustness of the control structure is verified by introducing noise to simulate the disturbance during anesthesia. The result shows that the control architecture could maintain the desired BIS level with disturbance rejection, meeting the clinical specifications.

In the future, we hope this scheme could be applied to actual clinical settings, not just simulation. For this purpose, it may be necessary to consider patient variability or equipment failure in the future. It can be the next step in our efforts.

## Declarations

The authors did not receive support from any organization for the submitted work.